\pdfoutput=1
\documentclass[runningheads]{llncs}
\usepackage[T1]{fontenc}
\usepackage{graphicx}
\usepackage{amsfonts,amssymb}
\usepackage{algorithm}
\usepackage{algorithmic}

\usepackage{bbm}
\usepackage{subcaption}
\usepackage{cite}
\begin{document}
\title{A Sample Reuse Strategy for Dynamic Influence Maximization Problem}
%
%
\author{Shaofeng Zhang\inst{1}
\and
Shengcai Liu\inst{2}
\and
Ke Tang\inst{1}
}

\authorrunning{S. ZHANG et al.}
%
\institute{Southern University of Science and Technology, Shenzhen 518055, China
\and
Centre for Frontier AI Research (CFAR), Agency for Science, Technology and Research (A*STAR), 138632, Singapore
}
\maketitle              
\begin{abstract}
Dynamic influence maximization problem (DIMP) aims to maintain a group of influential users within an evolving social network, so that the influence scope can be maximized at any given moment. A primary category of DIMP algorithms focuses on the renewal of reverse reachable (RR) sets, which is designed for static social network scenarios, to accelerate the estimation of influence spread. And the generation time of RR sets plays a crucial role in algorithm efficiency. However, their update approaches require sequential updates for each edge change, leading to considerable computational cost. 
In this paper, we propose a strategy for batch updating the changes in network edge weights to efficiently maintain RR sets. By calculating the probability that previous RR sets can be regenerated at the current moment, we retain those with a high probability. This method can effectively avoid the computational cost associated with updating and sampling these RR sets. Besides, we propose an resampling strategy that generates high-probability RR sets to make the final distribution of RR sets approximate to the sampling probability distribution under the current social network.
The experimental results indicate that our strategy is both scalable and efficient. On the one hand, compared to the previous update strategies, the running time of our strategy is insensitive to the number of changes in network weight; on the other hand, for various RR set-based algorithms, our strategy can reduce the running time while maintaining the solution quality that is essentially consistent with the static algorithms.

\keywords{Influence Maximization Problem  \and Dynamic Social Network \and Sample Reuse.}
\end{abstract}
\section{Introduction}
In recent years, with the rapid development of online social platforms, there is a huge potential for application analysis of social networks\cite{Kapoor_2018}. The influence maximization problem (IMP) stands as a representative problem within social network analysis. It aims at discovering a part of influential users, through which the spread of products or opinions could be maximized in the whole social network\cite{Kempe_2013}. The IMP has a wide range of applications, such as viral marketing\cite{LuWeiBonchi_2013}, election campaigns\cite{LitouIouliana_2018}, and fake news blocking\cite{ChenWenjie_2023}. However, in real-world scenarios, the relationships between users are constantly evolving, indicating that social networks are not static. The dynamic nature of social networks can affect the spread of influence\cite{DIM, Song_2017}. Every time the social network undergoes a change, the algorithms tailored for IMP must be restarted from scratch, resulting in huge computational overhead. Therefore, the dynamic influence maximization problem (DIMP) has been formulated, aiming to efficiently solve the IMP within dynamic social networks.

The primary existing approaches\cite{DIM,Yang_2017,Peng_2021} involve modifying RR set-based algorithms, which are efficient algorithms designed for IMP, adapting them to dynamic social networks. In RR set-based algorithms, the efficiency is heavily dependent on the generation
of RR sets. Therefore, they accelerate the evaluation of influence spread by updating previously generated RR sets, which can reduce the time costs associated with resampling.
However, their methods require sequential updates for each edge change in the network. For example, when the number of updated edge weights is \(a\), their methods would require \(a\) repetitive runs to process each change. In real-world scenarios, the dynamics of social networks are often captured through snapshots over a period of time, which can involve a substantial number of updates between two snapshots.
In such instances, the overhead of updating RR sets with their methods would exceed the cost of regenerating RR sets from scratch, making the updated algorithms less efficient than the static ones.

In this paper, we propose a new update strategy for RR set-based algorithms tailored to scenarios of network edge weight changes, which has the ability to simultaneously process multiple dynamic changes.
Our strategy can efficiently reduce the time costs of redundant sampling by preserving partial historically generated RR sets, which have a high probability of being regenerated at the current moment. This method does not require updating the RR sets for each updated weight, achieving the purpose of batch processing dynamic changes. 
Additionally, we propose a resampling strategy that generates new high-probability RR sets with previous rejected low-probability RR sets. This approach makes the final distribution of RR sets remains approximate to the probability distribution derived from sampling in the current social network.
The experiments indicate that our strategy is scalable. When the number of updated edge weights increases by \(9\) times, the running time of our algorithms only grow by \(0.3\) times, whereas the running time of the DynIM algorithm \cite{DIM} increases by more than \(60\) times. Moreover, our strategy is efficient, it can help RR set-based algorithms, IMM\cite{IMM} and SUBSIM\cite{SUBSIM}, achieve up to a \(19\%\) and \(12\%\) reduction in running time respectively.
\section{Preliminaries and Related Work}
\subsection{Problem Definition}
Dynamic influence maximization problem aims to maintain a group of influential users in dynamic environments, specifically within dynamic social networks, ensuring that their influence spread remain maximized at every moment. In this section, we will provide a formal definition of the DIMP.

{\bfseries Social Network:} The social network \( G = (V, E, P) \), where \( V = \{v_1, \dots, v_n\} \) represents the set of nodes, \( E \subseteq V \times V \) represents the edges between nodes, and \( P = \{p_{(u, v)} | (u, v) \in E\} \) represents the weights of the edges, reflecting the probability of influence propagation between nodes. Here, \( p_{(u, v)} \in (0,1] \) indicates the influence strength of node \( u \) on node \( v \).

{\bfseries Seed Set:} The seed set refers to the set of seed nodes \( S \subseteq V \) that are chosen to be activated in the initial state. 

{\bfseries Budget:} The budget \( k \) constrains the size of the seed set, meaning that at most \( k \) seed nodes can be selected.

{\bfseries Diffusion Model:} The diffusion model \( M \) captures the random process of information dissemination by the seed set in the social graph\cite{Li_2018}. The diffusion model adopted in this paper is the independent cascade (IC) model\cite{Kempe_2013}. In the IC model, each node has two states: activated and inactivated. The influence diffusion process unfolds in the following discrete steps:
\begin{itemize}
  \item In the initial step \( t =0 \), the nodes in the seed set  \( S \) are activated, while other nodes remain in inactive state.
  \item At step \( t \in [0, n] \), when node \( u \) is activated for the first time, it is considered contagious and has a single chance to activate each of its inactive out-neighboring nodes \( v \). The probability that node \( v \) gets activated is \( p_{(u,v)} \). Then, node \( u \) remains in an activate state but becomes non-contagious.
  \item The influence diffusion process terminates when no new nodes are activated, that is, when there are no more contagious nodes.
\end{itemize}

{\bfseries Influence Function:} Given a social network \( G = (V, E, P) \), a seed set \( S \subseteq V \), and a diffusion model \( M \), the influence function is defined as \( \sigma_{G, M}(S) \). It represents the expected number of nodes influenced (activated) by the seed set \(S\) when the influence diffusion process terminates\cite{Li_2018}.

{\bfseries Dynamic Social Network:} A dynamic social network is defined as a sequence of network snapshots evolving over time, \( \mathcal{G} = (G^0, G^1, \dots, G^T) \), where \( G^t = (V^t, E^t, P^t) \) is a snapshot of the network at time \( t \)\cite{Song_2017}. This study focuses on the updates of the network edge weight. Thus, the dynamic changes of the social network \( G^t \) are defined as \( \Delta G^t = (V, E, \Delta P^t) \). 

{\bfseries Dynamic Influence Maximization Problem:} Given a budget \( k \), the influence diffusion model \( M \), the social network snapshot at time \( t \) denoted as \( G^t \), and the dynamic changes \( \Delta G^t \), the dynamic influence maximization problem aims to find the seed set \( S^{t+1} \) in the social network snapshot \(G^{t+1} = G^t \cup \Delta G^t \) at time \( t+1 \), such that \( S^{t+1} \subseteq V \) and \( |S^{t+1}| \leq k \) to achieve influence maximization.
\begin{equation}
S^{t+1} = \mathop{\arg\max}_{S\in V,|S|<k} \sigma_{G^{t+1},M}(S)
\end{equation}
\subsection{Related Work}
The influence maximization problem was first proposed by Kempe\cite{Kempe_2013}, drawing inspiration from viral marketing strategies. They seeks to spread information throughout the social network using a word-of-mouth approach. There are two main difficulties in the influence maximization problem, one is that the influence of nodes can only be evaluated by Monte Carlo simulation rather than computed analytically, and the other is that the search space grows rapidly with the network size. 

Influence evaluation has been proved as \#P-hard problem\cite{Chen_2010}. Over recent years, many algorithms have been proved to estimate the influence of nodes. The simulation-based algorithms aim to reduce the overall evaluation cost by reducing unnecessary Monte Carlo sampling~\cite{LiLiu2023}, while still upholding theoretical guarantees\cite{CELF,UBLF}. Nonetheless, these methods still require huge computational time. The metrics-based algorithms utilize specially designed heuristic measures to approximate the influence of nodes \cite{ChenWei2009,Chen_2010},  but such methods tend to have lower accuracy. 
The sketch-based algorithms, particularly the RR set-based method\cite{RIS, SUBSIM,IMM}, are now widely adopted, because of their efficiency and theoretical guarantees. It employs a series of random generated RR sets \(\mathbbm{R}\) to concurrently approximate the influence of multiple nodes. Let \(R\) represent the RR set. The specific generation process in IC model is shown in Algorithm \ref{RR_gen}.
At this point, the influence of the seed set \(S\) is \(\frac{n}{|\mathbbm{R}|} \cdot |\Lambda_{\mathbbm{R}}(S)|\), where \(n\) is the number of nodes and \(\Lambda_{\mathbbm{R}}(S)|\) is the number of covered RR sets by \(S\)\cite{IMM}. Then we could use the greedy-based method to select influential seed seeds which ensures an approximation ratio of \( (1-\frac{1}{e}-\epsilon) \)\cite{Kempe_2013}. During the \(i^{th}\) iteration, with the current seed set denoted as \(S_{i-1}\), the greedy algorithm seeks a node \(v\) to maximize the marginal coverage \(|\Lambda_{\mathbbm{R}}(v|S_{i-1})|\), as illustrated in Equation \ref{coverage_gain}.
\begin{equation} \label{coverage_gain}
    \Lambda_{\mathbbm{R}}(v|S_{i-1}) = \Lambda_{\mathbbm{R}}(S_i\cup \{v\}) - \Lambda_{\mathbbm{R}}(S_{i-1})
\end{equation}
The \(\Lambda_{\mathbbm{R}}(v|S_{i-1})\) denotes the random RR sets within the set \(\mathbbm{R}\) that are covered by node \(v\) but remain uncovered by nodes in \(S_{i-1}\).
\begin{algorithm}
    \caption{RR-Set-Generation\cite{IMM}}
    \begin{algorithmic}[1] \label{RR_gen}
        \REQUIRE{social network $G$}.
        \ENSURE{RR set $R$}.
        \STATE RR set \(R\gets \emptyset\);
        \STATE Randomly select a node \(r \in V\) uniformly as the root node;
        \STATE Add node $r$ to queue $Q$ and RR set $R$;
        \STATE Set node $r$ as activated, and other nodes as inactivated;
        \WHILE{$Q$ has node}
            \STATE $v \gets Q.pop()$;
            \FOR {each node $u$ where $(u,v) \in E$}
                \IF{$u$ is inactivated \AND $rand() \leq p(u,v)$}
                    \STATE Add node $u$ to queue $Q$ and RR set $R$;
                    \STATE Mark node $u$ as activated;
                \ENDIF
            \ENDFOR
        \ENDWHILE
        \RETURN $R$;
    \end{algorithmic}
\end{algorithm}

However, in real-world scenarios, social networks are constantly evolving. There are some attempts to address the IMP in the dynamic social networks. Song pioneered a clear definition of the dynamic influence maximization problem (DIMP)\cite{Song_2017}, conceptualizing the dynamic social network as a sequence of static social networks. As such, DIMP can be seen as a series of static influence maximization problems. 
The historical evaluation information can then be used to help the algorithm efficiently update its estimation of node influence. 
For the metrics-based algorithms, updates can be applied to previous calculated metrics according to the dynamic changes in the network graph\cite{OhsakaNaoto_2015, WANG201910}. To reduce the update costs, local update strategies aim to heuristically limit the impact of network changes, thus preventing the need to update the influence of all nodes in social network\cite{WANG201910, Yalavarthi_2018}. For the sketch-based algorithms, especially the RR set-based algorithms, they tried to update the previously simulated RR sets, avoiding the computational cost associated with resampling\cite{DIM,Yang_2017,Peng_2021}. 
However, their algorithms need to resample for each changed edge and updating the affected RR sets, which can only process dynamic updates one by one to make the sampling probability of the RR sets matches the new social network. 
Therefore, their update strategy is effective primarily for scenarios with a small number of dynamic changes. As the number of dynamic updates increases, the cumulative time required for updates can surpass the time needed to regenerate RR sets from scratch, leading to poorer scalability of these algorithms. 
Besides, historical data can also assist in constructing seed sets more quickly. One approach directly updates the existing seed set. Specifically, it attempts to select new nodes from the social network to replace nodes in the existing seed set, aiming to maximize the influence gain\cite{Song_2017, Yalavarthi_2018}. This method can avoid the time cost of selecting seed sets from scratch. 
Additionally, when changes in the social network exhibit a certain pattern, we can leverage historical data to predict the network's change in the next time. Based on the predicted network dynamics, we can pre-select seed sets, thereby accelerating the solution selection\cite{Ashwini_2021, Zhang_2022}.

\section{Method}

\subsection{Sample Reuse With Importance Mixing}
In this section, we introduce a sample reuse algorithm that reuses samples based on their probability changes, aiming to avoid the computational cost caused by redundant sampling. RR set-based algorithms require the generation of a series of random RR sets, transforming the influence maximization problem into a maximum set coverage problem. In dynamic social networks, when the changes are relatively small, the random RR set generated at the previous time still have a high probability of being generated in the current moment. This paper takes the reverse influence sampling under the IC model as an example (as shown in Algorithm \ref{RR_gen}), treats the RR set obtained from sampling as a sample, and combines it with the Importance Mixing algorithm\cite{Sun_2009, Pourchot_2018} to reuse samples based on their probability changes.
\begin{figure}
    \centering
    \includegraphics[width=0.5 \linewidth]{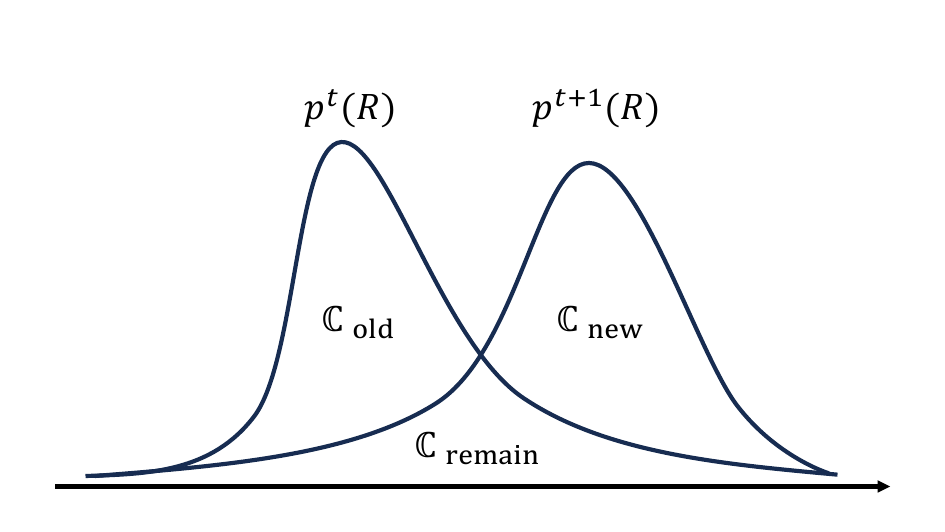}
    \caption{ Reverse reachable (RR) set \(R\) probability distribution in time \(t\) and \(t+1\)} \label{ImporMix}
\end{figure}

At the new moment \(t+1\), changes occur in the edges of the social network, that is, \( G^{t+1} = G^t \cup \Delta G^t \). We could use the Importance Mixing algorithm to update the series of RR sets \(\mathbbm{R}_{t}\) generated at the previous moment, ensuring that the updated RR sets satisfy the probability distribution of sampling under the new social network \(G^{t+1}\). 
Fig. \ref{ImporMix} depicts the evolution of the RR set's probability distribution. To make the previous RR sets (comprising both the \(\mathbbm{C}_{old}\) and \(\mathbbm{C}_{remain}\) parts) consistent with \(p^{t+1}(R)\), the Importance Mixing algorithm first rejects the \(\mathbbm{C}_{old}\) part and subsequently introduces the \(\mathbbm{C}_{new}\) part. In detail, the Importance Mixing approach reuses previous sampling RR sets in the following step:
\begin{itemize}
  \item \textbf{Remain Step:} For each previous RR set \(R_i\), which was generated in \(G^t\), the probability of reusing it is \(\min\{1, \frac{p^{t+1}(R_i)}{p^{t}(R_i)}\}\).
  \item \textbf{Sample Step:} Generate a new RR set  \(R_i'\) in the new social network \(G^{t+1}\). The probability of accepting it is \(\max\{0, 1 - \frac{p^{t}(R_i')}{p^{t+1}(R_i')}\}\).
\end{itemize}
\begin{figure}
    \centering
    \begin{subfigure}{0.4\textwidth}
        \includegraphics[width=\linewidth]{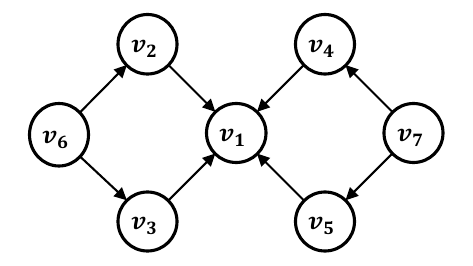}
        \caption{Graph}
    \end{subfigure}
    \hfill
    \begin{subfigure}{0.4\textwidth}
        \includegraphics[width=\linewidth]{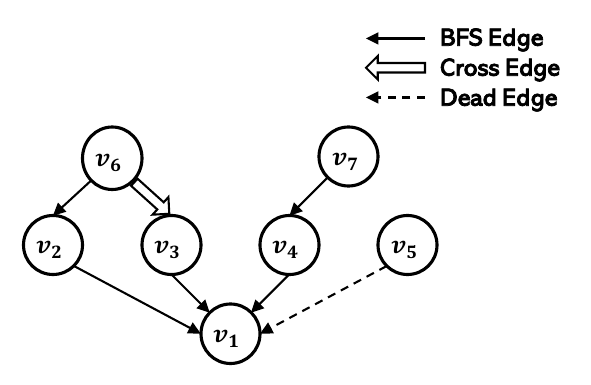}
        \caption{RR Set Sampling}
    \end{subfigure}
    \caption{An example of RR set sampling.} \label{example}
\end{figure}

The RR set-based algorithm for IMP only stored the activated nodes in the RR sets. The probability of the RR set is associated with the incoming edges of all activated nodes \(\{(u,v)| u\in R\}\). According to the definition in \cite{Yang_2017} and the procedure of Algorithm \ref{RR_gen}, we categorize these edges as BFS edge, Cross Edge, and Dead edge. For a specific edge \((u,v)\), the BFS edge signifies that node \(u\) was activated by node \(v\); the Cross edge indicates that when node \(v\) was activated, node \(u\) had already been activated; the Dead edge means node \(v\) tried to activate node \(u\), but was unsuccessful, as shown in Fig. \ref{example}. The probability of generating an RR set is given by Equation (\ref{prob}).
\begin{equation}
p(R)=\prod_{(u,v)\in E_{BFS}} p_{(u,v)} \cdot \prod_{(u,v)\in E_{Dead}} \left(1-p_{(u,v)}\right)
\label{prob} 
\end{equation}

However, due to the large number of Cross and Dead edges, storing and individually calculating probability for each RR set would lead to huge time and space costs. We develop a approximate method to efficiently calculate the update probabilities for RR sets, requiring storage of only the BFS edges for each RR set. When the social network undergoes changes, for each node \(u\), we calculate the probability of all its incoming edges being Dead edges, as shown in Equation (\ref{dead}), Where \(\lambda\) is a small value introduced to avoid division by zero.
\begin{equation}
p_{Dead} (u) = \prod_{(u,v)\in E} \frac{1-p^{t+1}_{(u,v)}+\lambda}{1-p^{t}_{(u,v)}+\lambda}
\label{dead} 
\end{equation}
Then, to compute the probability change for a RR set, we simply traverse each activate node in RR set and update the probabilities associated with the BFS edges. The resulting updated probability is shown in Equation \ref{new_prob}.
\begin{equation}
\frac{p^{t+1}(R)}{p^t(R)} \approx \prod_{(u,v)\in E_{BFS}} \left( \frac{p^{t+1}_{(u,v)}+\lambda}{p^{t}_{(u,v)}+\lambda} \cdot \frac{1-p^{t}_{(u,v)}+\lambda}{1-p^{t+1}_{(u,v)}+\lambda} \right) \cdot \prod_{u\in R} p_{Dead} (u)
\label{new_prob} 
\end{equation}
Through the reverse sampling process, as show in Algorithm \ref{RR_gen}, apart from the seed node, every activated node \(u\) has a unique parent node \(v\). Therefore, each activated node \(u\) corresponds to a unique BFS edge \((u, v)\). and the space occupied by storing BFS edges is consistent with that of storing activated nodes. 

\subsection{New Resampling Strategy}
In the \textbf{Sample Step} of Importance Mixing algorithm, RR sets are generated but may be rejected with a certain probability, which in turn wastes the time spent on sample generation. The purpose of the \textbf{Sample Step} is to generate RR sets with increased probability, as shown in the Fig. \ref{ImporMix} \(\mathbbm{C}_{new}\) part. However, compared to resampling from scratch, updating previously rejected samples is more likely to generate RR sets with increased probabilities. Specifically, when sample \(R\) is rejected and in the \(C_{old}\) part, it indicates that there is at least one edge \(e \in \{(u,v)|v\in R\}\) with a probability decrease.
\begin{itemize}
  \item If \((u,v)\) is a BFS edge, \(\frac{p^{t+1}_{(u,v)}}{p^{t}_{(u,v)}}<1\). Thus, when we remove this edge, it becomes a Dead edge and makes the RR set probability change to \(\frac{1-p^{t+1}_{(u,v)}}{1-p^{t}_{(u,v)}}>1\).
  \item If \((u,v)\) is a Dead edge, \(\frac{1-p^{t+1}_{(u,v)}}{1-p^{t}_{(u,v)}}<1\), adding the edge results in a RR set probability change of \(\frac{p^{t+1}_{(u,v)}}{p^{t}_{(u,v)}}>1\).
\end{itemize}

\begin{algorithm}
    \caption{Resample-RR-Set}
    \begin{algorithmic}[1] \label{RR_update}
        \REQUIRE{old RR set $R$, social network $G^{t+1}$}.
        \ENSURE{new RR set $R'$}.
        \STATE new RR set \(R'\gets \emptyset\);
        \STATE Select the root node \(v \in V\) in old RR set $R$;
        \STATE Add node $r$ to queue $Q$ and RR set $R'$;
        \STATE Set node $r$ as activated, and other nodes as inactivated;
        \WHILE{$Q$ has node}
            \STATE $v \gets Q.pop()$;
            \FOR {each node $u$ where $(u,v) \in E$}
                \IF{$u$ is activated}
                    \STATE \textbf{Continues};
                \ENDIF
                \IF{$p_{(u,v)}^t \not = p_{(u,v)}^{t+1}$}
                    \IF{$rand() \leq p_{(u,v)}^{t+1}$}
                        \STATE Add node $u$ to queue $Q$ and RR set $R'$;
                        \STATE Mark node $u$ as activated;
                    \ENDIF
                \ELSE
                    \IF{($u,v \in R$) \OR ($v\not \in R$ \AND $rand() \leq p_{(u,v)}^{t+1}$)}
                        \STATE Add node $u$ to queue $Q$ and RR set $R'$;
                        \STATE Mark node $u$ as activated;
                    \ENDIF
                \ENDIF
            \ENDFOR
        \ENDWHILE
        \RETURN $R'$;
    \end{algorithmic}
\end{algorithm}

Therefore, when the sample \(R\) is rejected in \textbf{Remain Step} of Importance Mixing algorithm, in \textbf{Sample Step}, we can resample the edges with changed probabilities, transitioning sample \(R\) from the \(\mathbb{C}_{old}\) part to the \(\mathbb{C}_{new}\) part. So, we propose a resampling strategy to generates new high-probability RR sets with previous rejected low-probability RR sets, as shown in Algorithm \ref{RR_update}.

Besides, we need to adjust the importance sampling part, as shown in Algorithm \ref{RIS-Generation-Importance-Mixing}. If the sample \(R\) is accepted in \textbf{Remain Step}, it suggests that the RR set \(R\) resides in the \(\mathbbm{C}_{remain}\), allowing us to move directly to the next iteration (Algorithm \ref{RIS-Generation-Importance-Mixing} Lines 3-5). However, if the sample \(R\) is rejected in \textbf{Remain Step}, indicating that its position is in the \(\mathbb{C}_{old}\) part, we then proceed to \textbf{Sample Step}. 
In this step, we resample with previous rejected low-probability RR set \(R\), as shown in Algorithm \ref{RR_update}, and accept it with the change probability to check whether the generated new RR set is in the \(\mathbb{C}_{new}\) part (Algorithm \ref{RIS-Generation-Importance-Mixing} Lines 6-11). Algorithm \ref{RIS-Generation-Importance-Mixing} Lines 16-21 ensure that the size of \(\mathbbm{R}_{t+1}\) is \( N_R \).

\begin{algorithm}
    \caption{RR-Sets-Generation-Importance-Mixing}\label{RIS-Generation-Importance-Mixing}
    \begin{algorithmic}[1]
        \REQUIRE{social network $G^{t+1}$, set of old random RR sets $\mathbbm{R}_{t}$, number of new RR sets $N_{R}$ }.
        \ENSURE{set of new random RR sets $\mathbbm{R}_{t+1}$}.
        \STATE $\mathbbm{R}_{t+1} \gets \emptyset$;
        \FOR{$i=1$ \textbf{to} $|\mathbbm{R}_{t}|$}
            \STATE $R_i \gets \mathbbm{R}_{t}[i]$;
            \IF{$\min (1,\frac{p^{t+1}(R_i)}{p^{t}(R_i)}) \geq rand() $ }
                \STATE $\mathbbm{R}_{t+1} \gets \mathbbm{R}_{t+1}.$append$(R_i)$;
            \ELSE
                \STATE $R_i' \gets$ \textbf{Resample-RR-Set($R_i$, $G^{t+1}$)}
                \IF{$\max (0,1-\frac{p^{t}(R_i')}{p^{t+1}(R_i')}) \geq rand() $ }
                    \STATE $\mathbbm{R}_{t+1} \gets \mathbbm{R}_{t+1}.$append$(R_i')$;
                \ENDIF
            \ENDIF
            \IF{$|\mathbbm{R}_{t+1}| \geq N_{R}$}
                \STATE \textbf{break};
            \ENDIF
        \ENDFOR
        \WHILE{$|\mathbbm{R}_{t+1}| > N_{R}$}
            \STATE remove a randomly chosen RR set $R$ in $\mathbbm{R}_{new}$;
        \ENDWHILE

        \WHILE{$|\mathbbm{R}_{t+1}| < N_{R}$}
            \STATE $R \gets$ Generate a new RR set in $G^{t+1}$;
            \STATE $\mathbbm{R}_{t+1} \gets \mathbbm{R}_{t+1}.$append$(R)$;
        \ENDWHILE
        \RETURN $\mathbbm{R}_{t+1}$;
    \end{algorithmic}
\end{algorithm}

\section{Experiments}
\subsection{Experimental Setting}
In this section, we will describe the details of the experimental settings. The experiments were executed on a Linux operating system, driven by an Intel(R) Xeon(R) Silver 4310 CPU @ 2.10GHz and equipped with 250GB of memory. Every algorithm we tested was crafted in C++ and compiled using g++, utilizing version 11.3.0 of the compiler.

{\bfseries Social Network Dataset:}
This paper utilizes two real-world network graph datasets for experimental evaluation, as summarized in Table \ref{datasets}.
The HepTh and HepPh datasets \cite{snapnets} are both citation graph, which are derived from the electronic literature platform, arXiv. Each node represents a paper, and edges signify citation relationships. 

\begin{table}
\caption{Datasets overview.}\label{datasets}
\centering
\begin{tabular}{|cccc|}
\hline
Dataset & Node Number & Edge Number & Type \\
\hline
HepTh & 27,770 & 352,807 & Directed\\
HepPh &  34,546 & 421,578 & Directed\\
\hline
\end{tabular}
\end{table}

{\bfseries Probability Setting:}
In this paper, we use the independent cascade diffusion model for influence spread. Since the weights in the social network are unknown, we adopt the widely-used weighted cascade (WC) model\cite{Kempe_2013}. Specifically, for each edge \(u,v\) in the network graph, the weight of the edge is \(p_{(u,v)} = \frac{1}{d_v^{in}}\), where \({d_v^{in}}\) represents the in-degree of node \(v\).

{\bfseries Algorithms:}
In this paper, we selected three algorithms, IMM, SUBSIM, and DynIM, as well as two algorithms proposed in this study, D-IMM and D-SUBSIM, for comparative experiments. 
\begin{itemize}
  \item \textbf{IMM\cite{IMM}:} The classic RR set-based algorithm for solving the IMP.
  \item \textbf{SUBSIM\cite{SUBSIM}:} The the state-of-the-art algorithm that uses RR sets to solve the IMP. 
  \item \textbf{DynIM\cite{DIM}:} The classic algorithm specifically designed to solve the DIMP.
  \item \textbf{D-IMM and D-SUBSIM:} Our proposed algorithms integrate the RR set-based algorithms, IMM and SUBSIM, respectively, with the sample reuse method introduced in this paper.
\end{itemize}

{\bfseries Parameter Setting:}
For the IMM, SUBSIM, D-IMM, and D-SUBSIM algorithms, the parameters are configured with values \( l = 1 \) and \( \epsilon = 0.1 \); for the DynIM algorithm, the parameter is set to \( \epsilon = 0.5 \).
In the experimental setup, the budget for the seed set \(k\) is fixed at \(50\). For the seed sets obtained by different algorithms, we use the Monte Carlo simulation method to estimate the influence spread, with the number of simulations \(r=10,000\). Due to the randomness inherent in the algorithms, each experiment is repeated \(10\) times, and the running time and influence are averaged~\cite{LiuTL020}. Note that when evaluating running time, since every algorithm needs to compute on in the updated graph, we excluded the time required to load and update the social network.

\subsection{Edge Weight Update Analysis}
\begin{figure}
    \centering
    \begin{subfigure}{0.33\textwidth}
        \includegraphics[width=\linewidth]{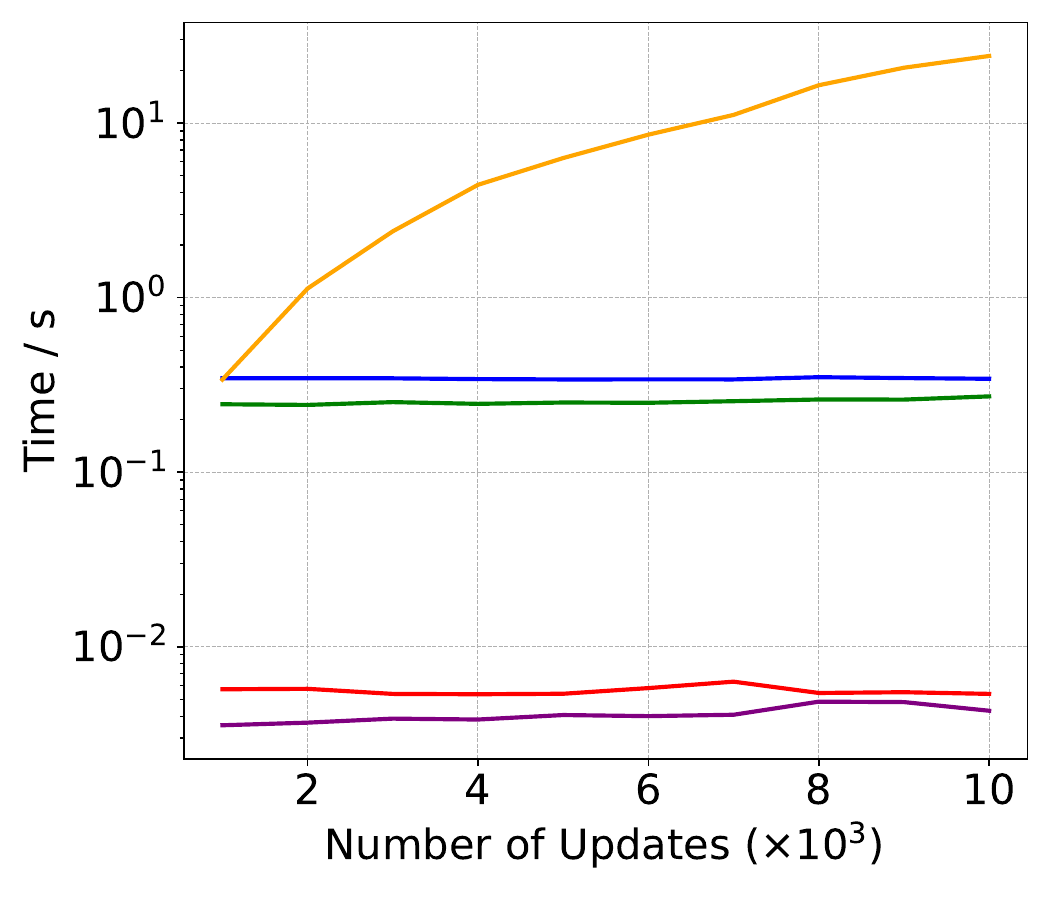}
        \caption{HepTh}\label{HepTh_times}
    \end{subfigure}
    \hfill
    \begin{subfigure}{0.47\textwidth}
        \includegraphics[width=\linewidth]{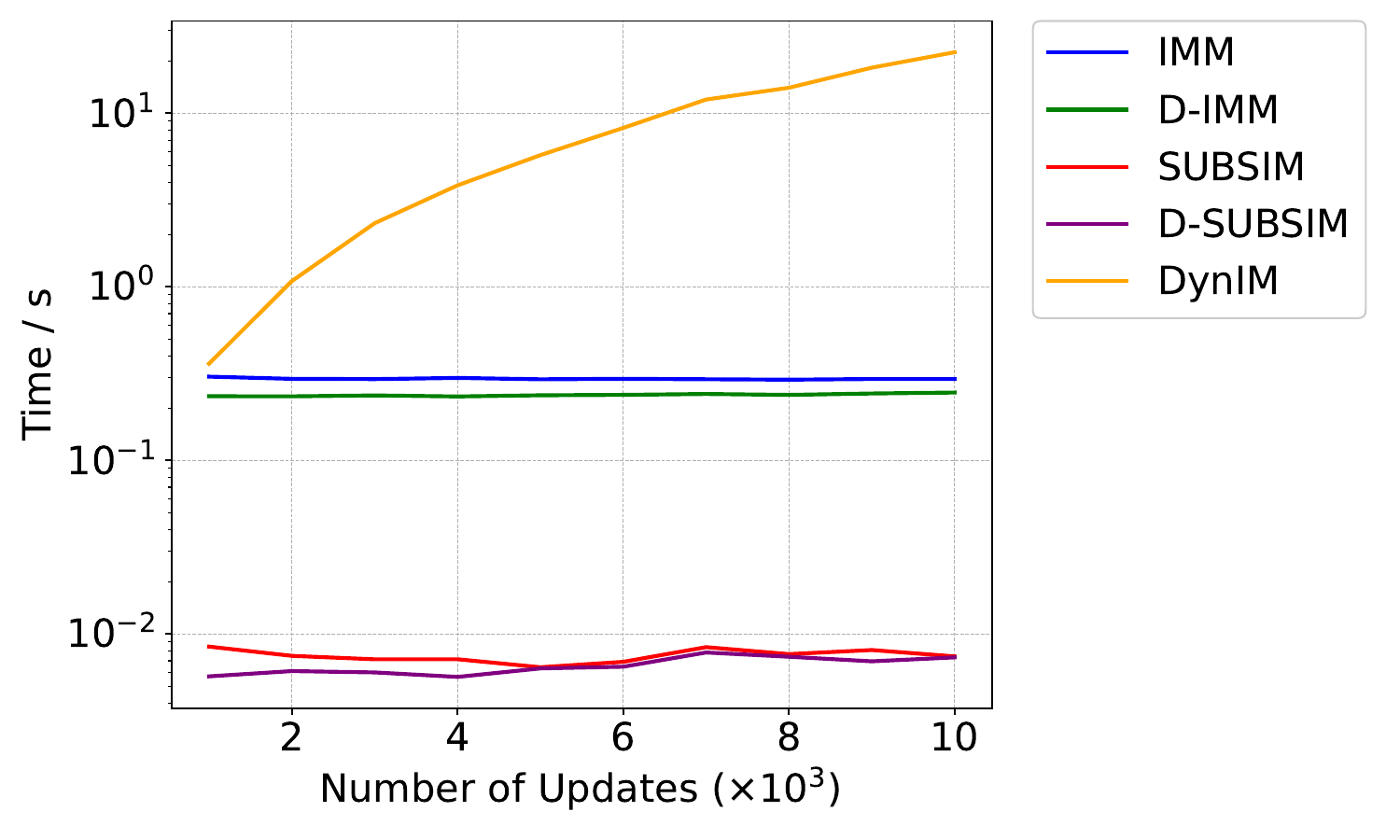}
        \caption{HepPh}\label{HepPh_times}
    \end{subfigure}
    \caption{The running time of the algorithms under different numbers of edge weight updates.} \label{algo_times}
\end{figure}

\begin{figure}
    \centering
    \begin{subfigure}{0.33\textwidth}
        \includegraphics[width=\linewidth]{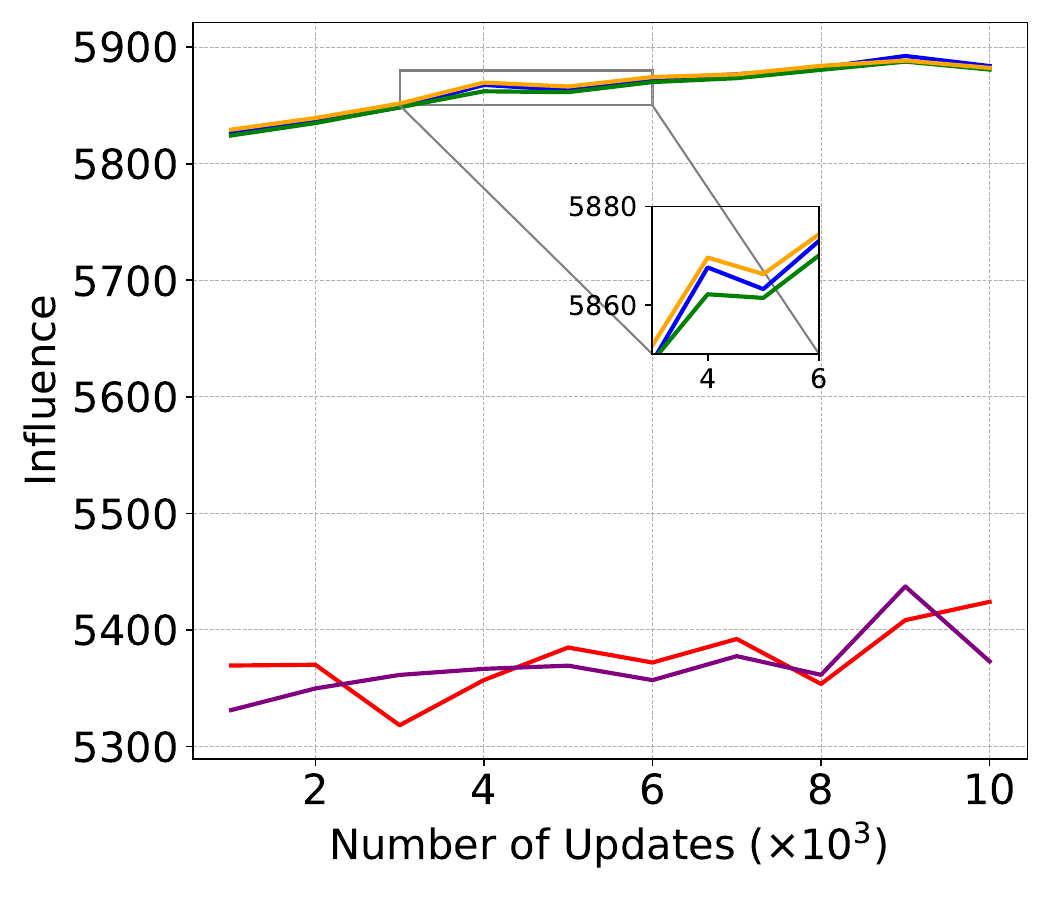}
        \caption{HepTh}\label{HepTh_inf}
    \end{subfigure}
    \hfill
    \begin{subfigure}{0.47\textwidth}
        \includegraphics[width=\linewidth]{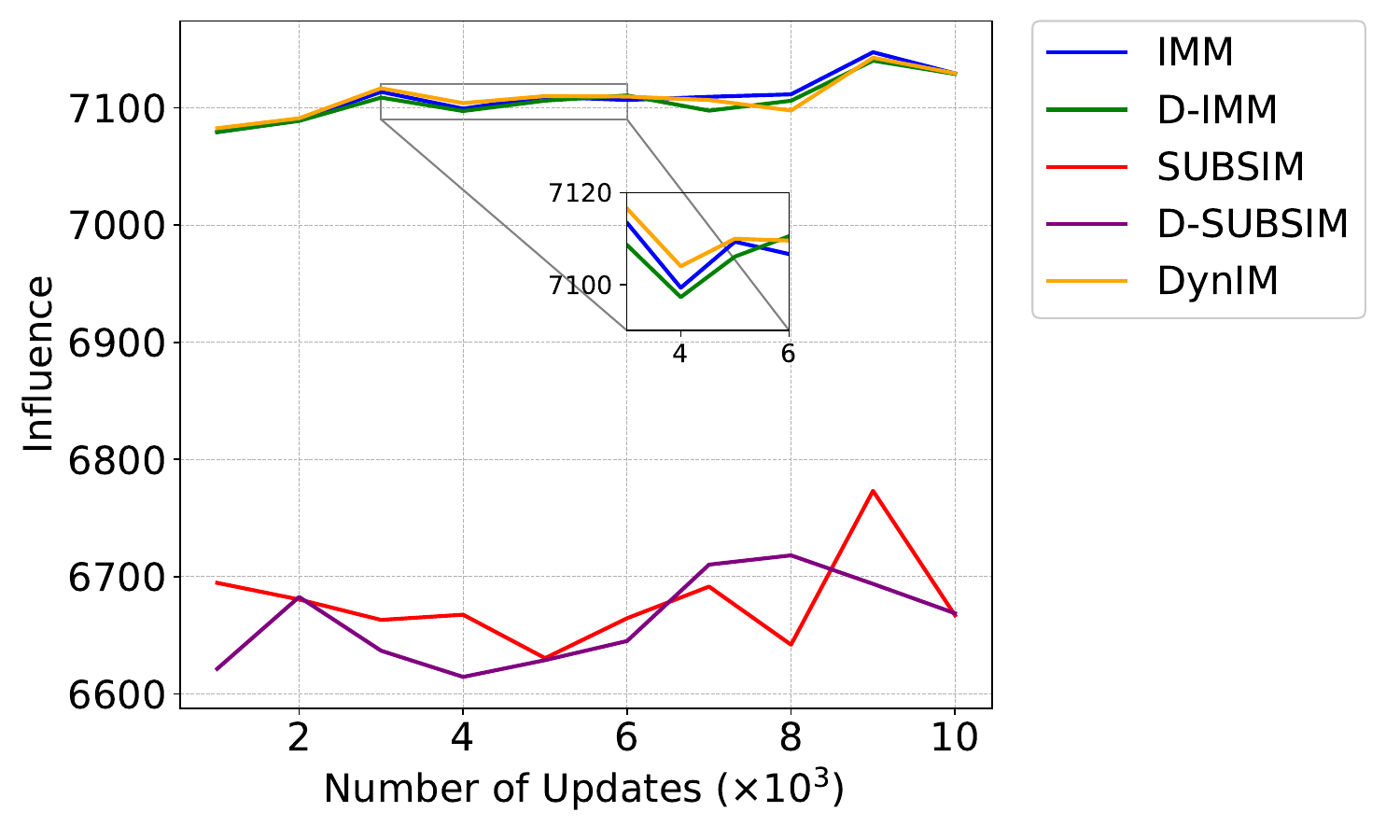}
        \caption{HepPh}\label{HepPh_inf}
    \end{subfigure}
    \caption{The influence spread of the algorithms under different numbers of edge weight updates.} \label{algo_inf}
\end{figure}

This experiment aims to examine the algorithms' efficiency and scalability with respect to the number of dynamic updates. We focus on scenarios with edge weight changes to isolate the effect of network size variations on the algorithms' running time.
The experiment involves only two snapshots of the social network, that is, the social network undergoes one change involving multiple dynamic updates. 
Specifically, following the settings in \cite{DIM}, each dynamic update represents randomly selecting an edge \((u, v)\) and change its probability to \(p^{t} \times 2\) or \(p^{t} / 2 \). For each social network, we varied the number of dynamic updates to evaluate the algorithm's performance under various dynamic scenarios in terms of running time (as shown in Fig. \ref{algo_times}) and the solution quality (as shown in Fig. \ref{algo_inf}).

For the datasets HepTh and HepPh, we varied the number of updated social network edge weights within the range of \([10^3, 10^4]\) and conducted comparative experiments across all the five algorithms. As shown in Fig. \ref{HepTh_times} and Fig. \ref{HepPh_times}, 
the average running time of the DynIM algorithm for processing \(10^4\) updated edges is \(71.5\) times and \(62.3\) times, respectively, of the time taken for \(10^3\) updated edges. In contrast, our algorithm's running time (D-IMM and D-SUBSIM) for \(10^4\) updated edges is both merely \(1.3\) times that of \(10^3\) updated edges for two datasets. 
Therefore, compared to the DynIM algorithm, our method's running time, D-IMM and D-SUBSIM, is less sensitive to the number of updates. 
Moreover, for the datasets HepTh and HepPh, the time taken by the DynIM algorithm to update its RR sets is significantly higher than that of the other four algorithms.
While the D-IMM algorithm achieves a \(26.2\%\) and \(19.3\%\) reduction in average running time compared to the IMM algorithm, and similarly, the D-SUBSIM algorithm exhibits a \(26.6\%\) to \(12.5\%\) decrease in average running time when contrasted with the SUBSIM algorithm. Therefore, for the scenarios of network edge weight change, our method can effectively accelerate the IMM and SUBSIM algorithms.

Figure \ref{algo_inf} indicates that the solution quality of our algorithm is comparable to that of the static algorithms. For the two datasets, the average influence spread of the seed sets solved by the D-IMM algorithm shows a difference of less than \(0.5\%\) compared to the IMM algorithm. Besides, the average influence spread of the D-SUMSIM algorithm differs by no more than \(0.3\%\) from that of the SUMSIM algorithm, which is still within an acceptable range. Therefore, our dynamic algorithms, D-IMM and D-SUBSIM, do not show a significant decline in solution quality compared to the static algorithms.

\section{Conclusion}
In this paper, we propose a new update strategy for RR set-based algorithms which can efficiently handle batch updates in dynamic network. Specifically, when changes occur in the social network, it reuses historical RR sets, which still have a high probability of being regenerated at the current time, to avoid the computational cost caused by redundant sampling. Additionally, we design an efficient resampling strategy to generate new high-probability RR sets with previous rejected low-probability RR sets, which makes the final distribution of RR sets remain approximate to the probability distribution derived from sampling in the new social network. 
The experimental results demonstrate that our algorithm exhibits better scalability compared to the previous update strategy. Besides, our strategy can effectively reduce the running time of RR set-based algorithms. 
While our algorithm is efficient, the resampling strategy does not provide a theoretical guarantee that the final probability distribution of the RR sets is consist with the one derived from sampling in the new social network. In the future, we will attempt to improve the algorithm and propose a sampling strategy with stronger theoretical assurances.
%
%
%
\bibliographystyle{splncs04}
\bibliography{Ref}

\end{document}